\documentclass[twocolumn, showpacs, superscriptaddress,preprintnumbers, amsmath, epsfig, floatfix, prl]{revtex4}

\usepackage{graphicx}
\usepackage{dcolumn}
\usepackage{bm}
\usepackage{ulem}

\usepackage[utf8]{inputenc}
\usepackage[english]{babel}
\usepackage{keyval}
\graphicspath{{figs/}{figs:}{\figs}}
\setkeys{Gin}{width=0.90\columnwidth}

\usepackage[usenames,dvipsnames]{color}
\usepackage[]{ulem}

\newcommand{\comment}[1]{\textcolor{red}{#1}}
\renewcommand{\comment}[1]{\relax}

\newcommand{\todelete}[1]{\textcolor{green}{\sout{#1}}}
\renewcommand{\todelete}[1]{\relax}

\begin{document}

\title{Emergent antiferromagnetism of YTiO$_3$ in YTiO$_3$-CaTiO$_3$ superlattices}
\date{\today}
\author{P. Pal}
\affiliation{Department of Physics, Indian Institute of Technology Kharagpur, Kharagpur-721302, India}
\author{X. Liu}
\affiliation{Department of Physics and Astronomy, Rutgers University, Piscataway, New Jersey 08854, USA}
\author{M. Kareev}
\affiliation{Department of Physics and Astronomy, Rutgers University, Piscataway, New Jersey 08854, USA}
\author{D. Choudhury}
\email{debraj@phy.iitkgp.ac.in}
\affiliation{Department of Physics, Indian Institute of Technology Kharagpur, Kharagpur-721302, India}
\author{J. Chakhalian}
\affiliation{Department of Physics and Astronomy, Rutgers University, Piscataway, New Jersey 08854, USA}

\begin{abstract}
\noindent Transport and magnetoresistance measurements are performed on metallic, high-carrier density YTiO$_3$-CaTiO$_3$ superlattices as a probe towards the investigation of an emergent magnetic order of YTiO$_3$. On varying the thickness of YTiO$_3$ while keeping the CaTiO$_3$ layer thickness constant in the superlattices, a low-temperature upturn in sheet-resistance, a non-Fermi liquid-like charge transport and positive magnetoresistance are observed. Analyses of the origin of such effects suggest that a unique antiferromagnetic order is realized in the ultra-thin, epitaxially strained YTiO$_3$ layers, which corroborates well with some recent theoretical predictions in this regard.

\end{abstract}

\maketitle
\begin{figure}[t]
\vspace*{-0.23 in}
\hspace*{-0.22 in}\scalebox{1.2}{\includegraphics{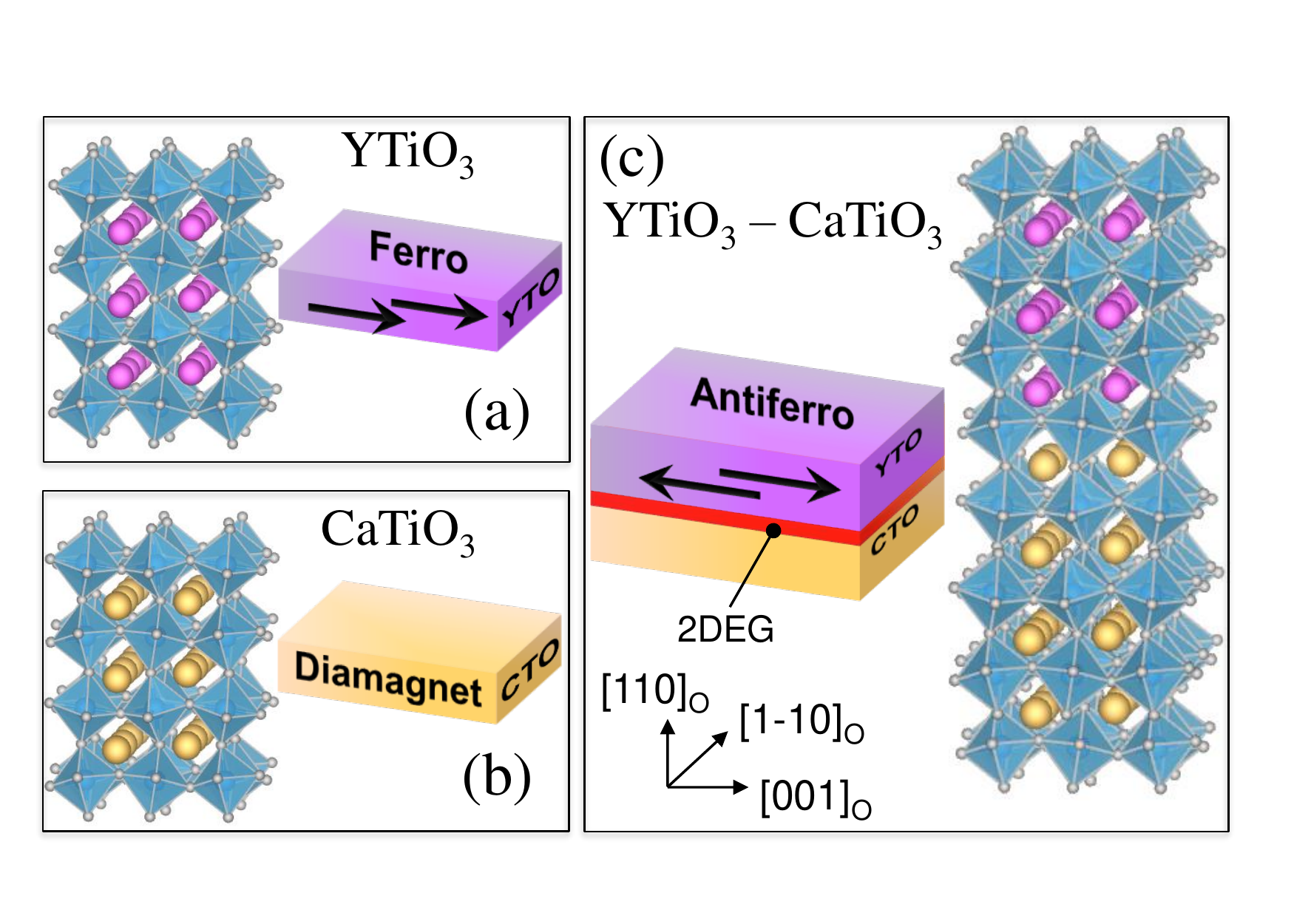}}
\vspace*{-0.4 in}\caption{(Color online) Schematic representations of the orthorhombic structures of (a) YTiO$_3$ (which is ferromagnetic in bulk form), (b) CaTiO$_3$ (which is diamagnetic in bulk form), and (c) of the emergence of antiferromagnetism of YTiO$_3$ in the thin-film superlattice of YTiO$_3$ - CaTiO$_3$.}\label{Structure}
\end{figure}

Tunability of the magnetic ground state of transition metal oxides through its coupling to the orbital and structural degrees of freedom is an extremely interesting field of research and leads to plethora of functional properties \cite{YTokura2000,JBGoodenough1955,DIKhomskii1982}. Recently, a lot of attention has been focussed on investigating the spin and orbital orders of the rare-earth titanate $\it{R}$TiO$_3$ (where $\it{R}$ represents a trivalent rare-earth ion) compounds as they exhibit strongly intertwined structural, spin and orbital orders \cite{MBibes2017,BKeimer2009,MImada2004,SOkamoto2002}. $\it{R}$TiO$_3$ compounds comprise of Ti$^{3+}$ (3$\it{d}^{\rm1}$) magnetic ions and, owing to the large on-site Coulomb energy, are Mott-Hubbard insulators \cite{MImada2004}. Due to  small tolerance factor, $\it{R}$TiO$_3$ compounds adopt an orthorhombic $\it{P}$bnm structure characterized by large tilts between their TiO$_6$ octahedral cages (termed as GdFeO$_3$-type distortion) (shown in Fig. \ref{Structure}). The GdFeO$_3$-type distortion of $\it{R}$TiO$_3$ compounds can be parameterized using the deviation of the $\it{Ti}$-$\it{O}$-$\it{Ti}$ bond angle from the undistorted value of 180$^o$ and it's magnitude depends critically on the size of the particular $\it{R}$ ion in $\it{R}$TiO$_3$ \cite{MImada2004}. Going from the smaller Y ion towards the larger La ion, the $\it{Ti}$-$\it{O}$-$\it{Ti}$ bond angle measured along the crystallographic $\it{c}$-axis changes from $\sim$140$^o$ (large GdFeO$_3$-type distortion) to $\sim$156$^o$ (smaller GdFeO$_3$-type distortion) \cite{MImada2004}. The magnetic and orbital orderings of $\it{R}$TiO$_3$ compounds have been found to depend critically on the magnitude of the GdFeO$_3$-type distortion, which, along with the associated lattice distortions, leads to an effective Jahn-Teller distortion on the Ti site and also induces $\it{t}_{2g}$-$\it{e}_g$ orbital mixing \cite{MBibes2017,MImada2004}. An increase (decrease) of the GdFeO$_3$-type distortion usually favours ferromagnetic (antiferromagnetic) coupling between the Ti$^{3+}$ (3$\it{d}^{\rm1}$) moments \cite{BKeimer2009}. We note that a recent experimental result, surprisingly, seems to suggest that the magnetic ground-state of GdTiO$_3$ thin-films is largely independent of the TiO$_6$ octahedral-tilts \cite{SDWilson2017}.

\indent A key theoretical prediction in regards to $\it{R}$TiO$_3$ compounds is that a novel antiferromagnetic (A-type) phase can be stabilized in YTiO$_3$ (YTO) by carefully tuning its GdFeO$_3$-type distortion \cite{MImada2004}. AFM materials, unlike their FM counterparts, have become the focus of interest for future spintronics devices \cite{TJungwirth2010}. YTO, in bulk-form, is ferromagnetic (FM) with a Curie temperature of 30 K. Most likely due to the difficulty of obtaining bulk samples with such fine-tuned lattice distortions, experimental efforts, like in Y$_{1-\it{x}}$La$_{\it{x}}$TiO$_3$ bulk solid solutions, have not been able to detect the predicted A-type antiferromagnetic (AFM) phase \cite{YTokura1995,DAMacLean1982}. Using first-principles calculations, epitaxially-strained YTO thin-films grown on LaAlO$_3$ (LAO) (001) substrate has been predicted to host the A-type AFM phase of YTO \cite{SDong2015,SDong2013}.

However, due to the associated large epitaxial strain ($\sim$ 5$\%$), growth of epitaxially strained YTO film on LAO (001) substrate is difficult to be realized experimentally \cite{SDong2015}. Thin-film heterostructure superlattices provide an effective platform to controllably tune the magnitude of the GdFeO$_3$-type lattice distortions by tuning the thickness of the individual constituent layers in these superlattices \cite{SStemmer2013} and lead to many interesting physical phenomena, like metal-insulator transition \cite{SStemmer2012,SStemmer2014}, quantum-critical fluctuations, non-Fermi liquid behavior of charge-carriers \cite{SStemmerNC2014}, and interface-induced magnetism \cite{SStemmerPRB2013}.

\indent We have recently succeeded in obtaining a non-STO based two-dimensional electron-gas (2DEG) in the YTO - CaTiO$_3$ (CTO) superlattices with very high sheet carrier densities ($\simeq$ 10$^{\rm{14}}$ cm$^{\rm{-2}}$) \cite{JChakhalian2015}. Interestingly, the heterostructure of YTO with STO is found to be insulating \cite{JChakhalian2013}. CTO crystallizes in orthorhombic $\it{P}$bnm structure, and, unlike YTO, contains Ti$^{\rm{4+}}$ (3$\it{d}^{\rm0}$) ions, possesses smaller GdFeO$_3$ distortion ($\it{Ti}$-$\it{O}$-$\it{Ti}$ bond angle is $\sim$157$^o$ measured along the crystallographic $\it{c}$-axis) and is a diamagnetic band-insulator \cite{JChakhalian2016}. Magnetoresistance phenomenon has been the focus of attention as an effective and sensitive tool to probe AFM ordering even in AFM multidomain materials \cite{SGeprags2018,GEWBauer2013,STakada1973}. In this letter, we report on investigations of the transport and magnetoresistance (MR) properties carried out on YTO-CTO superlattices containing different number of pseudocubic layers of YTO (varying GdFeO$_3$ distortions) towards the realization of a novel AFM phase of YTO in YTO-CTO heterostructure.

\indent Epitaxial [$\it{m}$YTO - 6CTO]$_4$ superlattices [Here $\it{m}$ and 6 stand for the number of pseudocubic unit-cells of YTO and CTO, respectively, within a heterostructure-unit and 4 represents the number of stacked units in the superlattice] (for $\it{m}$=3 and 6) and [3YTO-6CTO]$_5$ were grown on NdGaO$_3$ (NGO) (110)$_{\rm{O}}$ and DyScO$_3$ (DSO) (110)$_{\rm{O}}$ [orthorhombic notation] substrates, respectively, with abrupt interfaces using pulsed laser deposition technique. The films were characterized using in-situ reflection high-energy electron diffraction, x-ray diffraction and x-ray reflectivity measurements. Transport and magnetoresistance measurements (with both out-of-plane and in-planar magnetic field applied perpendicular to current direction) were carried out in van-der Pauw geometry using ohmic contacts in a physical property measurement system. Details of the growth conditions, structure and methodologies are reported in Ref. \cite{JChakhalian2015}. For YTO-CTO superlattices, grown on NGO (110)$_{\rm{O}}$, and also for DyScO$_3$ (DSO) (110)$_{\rm{O}}$-oriented substrates, the growth direction of the films is also along the orthorhombic (110)$_{\rm{O}}$ direction of YTO and CTO, in line with the subsrates. The orthogonal axes, (1-10)$_{\rm{O}}$ and (001)$_{\rm{O}}$, thus, charaterize the in-plane lattice structure of the individual thin-film layers as represented in Fig. \ref{Structure}.

\begin{table}[h]
  \centering
  \caption{Lattice parameters of orthorhombic YTiO$_3$ (YTO), CaTiO$_3$ (CTO), NdGaO$_3$ (NGO) and DyScO$_3$ (DSO) compounds. The percentage differences taking YTO as reference are indicated within brackets.}
  \label{3tablatparameter}
  \vspace*{0.5cm}
\begin{tabular}{|c|c|c|c|}
  \hline
   & Along (1-10)$_{\rm{O}}$ $(\AA)$ & Along (001)$_{\rm{O}}$ $(\AA)$ \\
  \hline
  YTO & 7.779 & 7.611  \\
  CTO & 7.652 (-1.63$\%$) & 7.640 (+0.38$\%$) \\
  NGO & 7.726 (-0.68$\%$) & 7.708 (+1.27$\%$) \\
  DSO & 7.905 (+1.62$\%$) & 7.913 (+3.96$\%$) \\
  \hline
\end{tabular}

\end{table}

\begin{figure}[h]
\vspace*{0.0 in}
\hspace*{-0.22 in}\scalebox{1.43}{\includegraphics{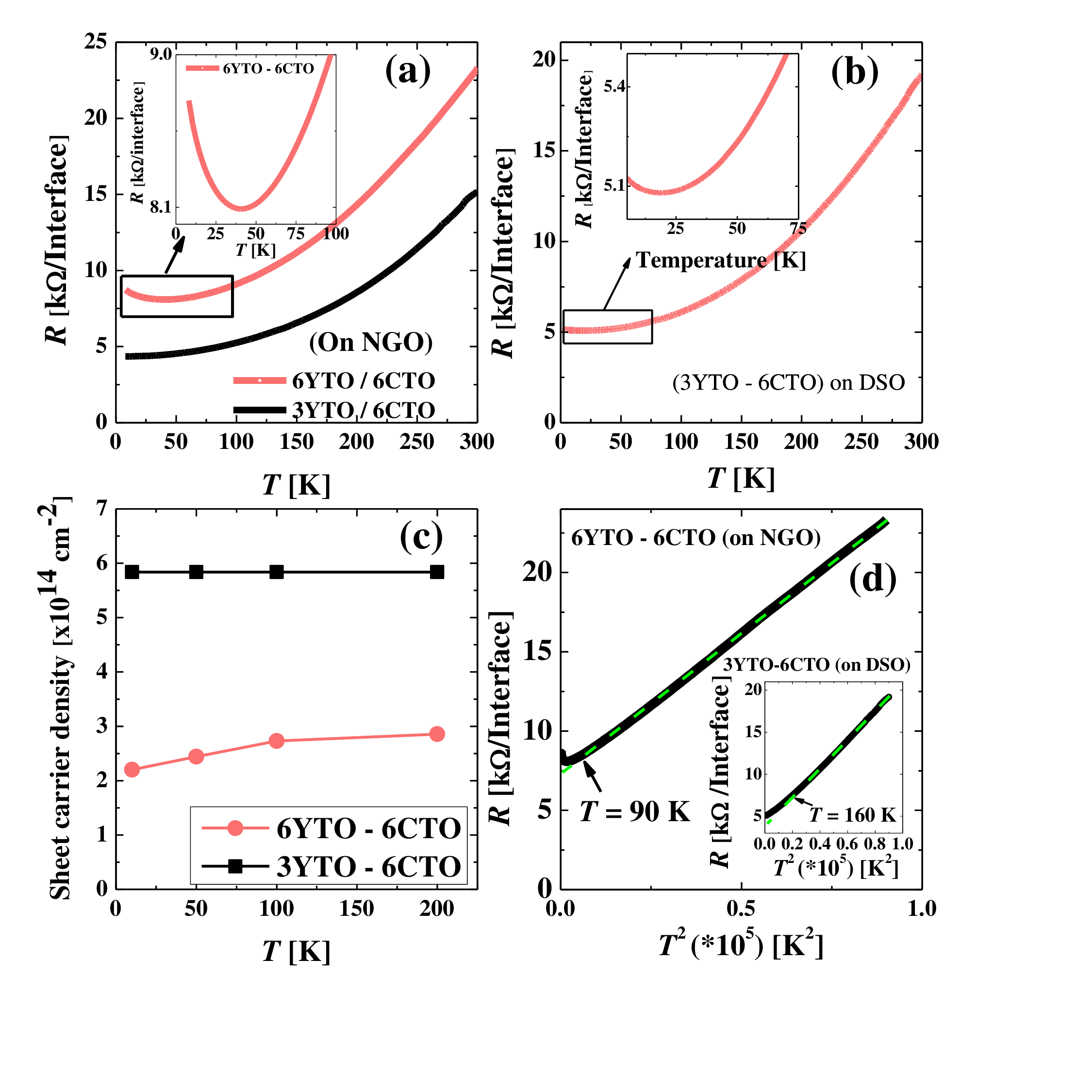}}
\vspace*{-0.6 in}\caption{(Color online)(a) Temperature ($\it{T}$)- dependence of sheet-resistance ($\it{R}$) of $\it{m}$YTiO$_3$-6CaTiO$_3$ superlattices, for $\it{m}$=3 and 6 on NdGaO$_3$ (NGO) substrate. Inset to (a) shows an expanded view of the low-$\it{T}$ region for $\it{m}$=6. (b) $\it{T}$-dependence of $\it{R}$ for $\it{m}$=3 superlattice on DyScO$_3$ (DSO) substrate. Inset to (b) shows an expanded view of low-$\it{T}$ region of corresponding $\it{R}$-data. (c) $\it{T}$-dependence of the sheet-carrier densities, estimated from Hall data, for $\it{m}$=3 and 6 superlattices grown on NGO. (d) $\it{R}$ for $\it{m}$=6 superlattice plotted against $\it{T}^{\rm{2}}$. The inset shows the same for $\it{m}$=3 superlattice grown on DyScO$_3$. The thick solid straight lines in these plots are guide to eyes.}\label{ResT}
\end{figure}

\indent The temperature-dependence of the sheet-resistance ($\it{R}$) data measured on the $\it{m}$=3 and 6 superlattices, grown on NGO, are shown in Fig. \ref{ResT}(a). The $\it{m}$=3 superlattice exhibits metallic-like transport down to the lowest measurement temperature. The $\it{m}$=6 superlattice, however, exhibits metallic conductivity till $\simeq$40 K and an upturn in $\it{R}$-value below this temperature. Using DSO (110)$_{\rm{O}}$-oriented substrate, which induces larger tensile strain on YTO along (001)$_{\rm{O}}$ than NGO (110)$_{\rm{O}}$, even $\it{m}$=3 superlattice is observed to exhibit a low-temperature upturn in $\it{R}$ data, as seen in Fig. \ref{ResT}(b). The Hall-resistance data for the YTO-CTO superlattices which were found to be negative and linearly dependent on applied magnetic field, suggest electrons as a single species of charge carriers in these superlattices. The presence of electrons as the carrier type also suggests that the carriers reside on the CTO side of the YTO-CTO heterointerface. The estimated sheet carrier densities for $\it{m}$=3 and 6 superlattices, estimated from the Hall-data and shown in Fig. \ref{ResT}(c), were found to be nearly independent of temperature, suggesting no significant change of electronic structure till the lowest temperature. As emphasized in Ref. \cite{JChakhalian2015}, individual YTO and CTO thin-film layers, grown under identical growth directions on NGO, exhibit strongly insulating behavior and possesses resistance values beyond the range of our measurement apparatus, thereby proving that the metallic conductivity arises from the YTO-CTO heterointerface and not from the individual layers. Interestingly, the sheet carrier density of the $\it{m}$=6 superlattice ($\simeq$3$\times$10$^{14}$ cm$^{-2}$) is slightly smaller than the $\it{m}$=3 superlattice ($\simeq$6$\times$10$^{14}$ cm$^{-2}$). This is opposite to what should be expected if the conductivity is due to electron doping in CTO through Y-ion doping and supports the origin of 2DEG to be related to interfacial charge-transfer from YTO into CTO, as expected from their band-alignments \cite{CGVandeWalle2014}.

\indent To investigate the nature of scattering mechanisms of charge-carriers in these superlattices, the temperature ($\it{T}$) exponents of $\it{R}$ were investigated. A Fermi-liquid-like $\it{T}^{\rm{2}}$ dependence of $\it{R}$, as expected from electron-electron scattering mechanism, is observed above $\simeq$90 K, $\simeq$160 K and $\simeq$200 K for $\it{m}$=6 superlattice on NGO (see Fig. \ref{ResT}(d)), $\it{m}$=3 superlattice on DSO (inset to Fig. \ref{ResT}(d)) and $\it{m}$=3 superlattice on NGO (not shown for brevity), respectively. Below these temperatures, the $\it{R}$ values are found to be higher than the value estimated from the simple $\it{T}^2$ dependence at higher temperatures, which suggests additional scattering channels for the conduction electrons at lower temperatures. Interestingly, the $\it{m}$=3 superlattice on NGO, which exhibits non-Fermi liquid like charge transport till the highest temperature ($\simeq$200 K) exhibits no upturn in low-temperature $\it{R}$-values.

\begin{figure}[t]
\vspace*{0.0 in}
\hspace*{-0.2 in}\scalebox{1.43}{\includegraphics{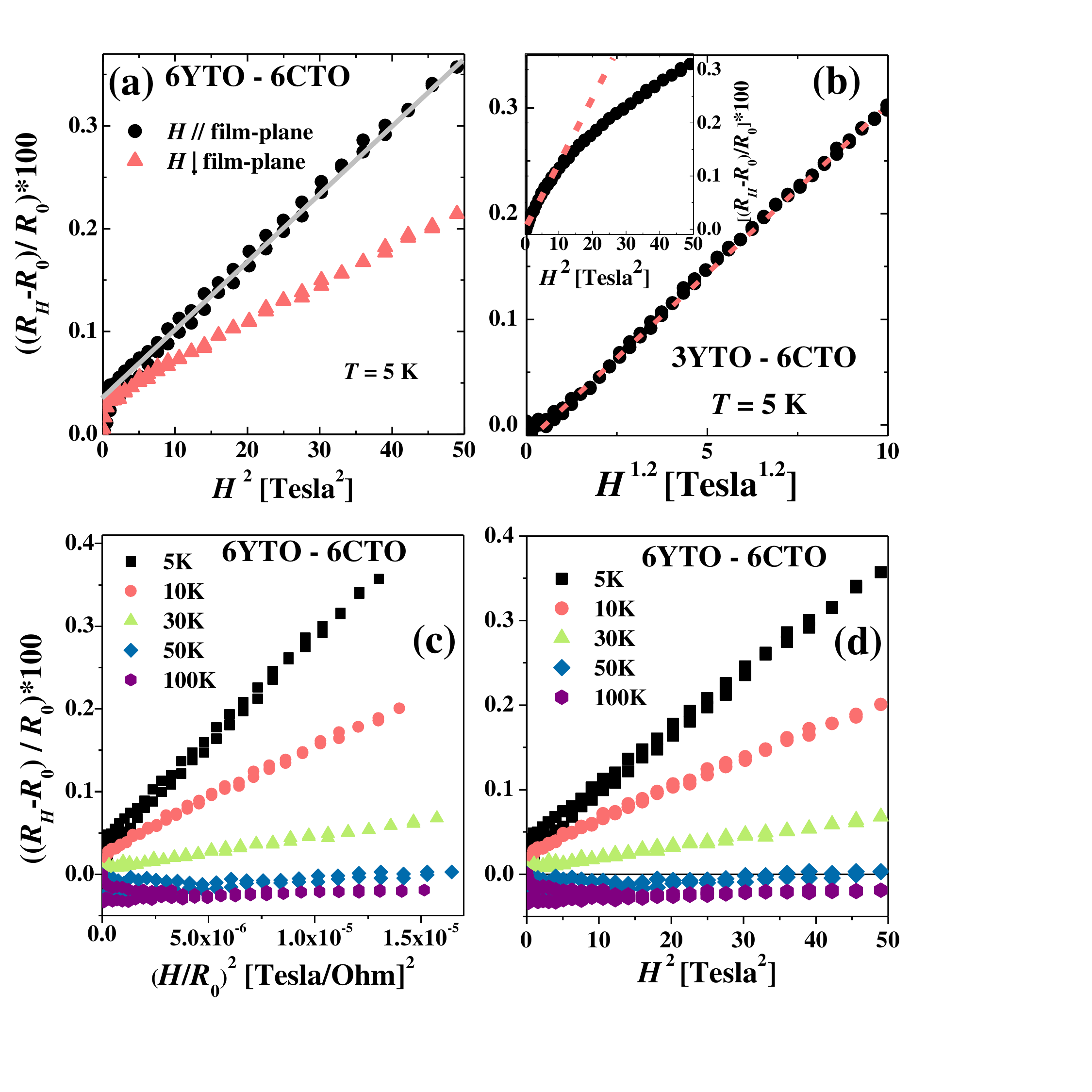}}
\vspace*{-0.6 in}\caption{(Color online)(a) The magnetoresistance (MR) of $\it{m}$=6 superlattice at $\it{T}$= 5 K calculated from $\it{R}_{\rm{H}}$-data (sheet-resistance in presence of magnetic field $\it{H}$) plotted against $\it{H}^{\rm{2}}$ for $\it{H}$ applied perpendicular to and in-thin-film-plane. (b) The MR data at $\it{T}$= 5 K of $\it{m}$=3 superlattice plotted with $\it{H}^{\rm{1.2}}$. The inset to (b) shows the the corresponding MR plotted against $\it{H}^{\rm{2}}$. The short-dashed straight lines are guide to eyes. (c) The MR isotherms against ${(\frac{H}{R_{\rm{0}}})}^{\rm{2}}$ for $\it{m}$=6 showing that corresponding $\it{T}$-dependent MR does not scale according to Kohler's rule. (d) The MR isotherms of $\it{m}$=6 superlattice plotted against $\it{H}^{\rm{2}}$ showing a sign-change of MR between $\it{T}$=30 K and 50 K.}\label{ResH}
\end{figure}

\indent To investigate the origin of low-temperature upturn in $\it{R}$-values, $\it{T}$- and magnetic-field ($\it{H}$)- dependent MR measurements were performed on the superlattices grown on NGO (MR measurements could not be performed on superlattices on DSO substrate because of the large magnetic anisotropy of DSO). Interestingly, a positive MR is observed for both $\it{m}$=3 and 6 superlattices (as shown in Fig. \ref{ResH}(a) and (b)) (for $\it{H}$ applied both perpendicular to and in the thin-film plane). We note here, that a similar low-temperature upturn in $\it{R}$-values was observed for SmTiO$_3$ (SmTO)-SrTiO$_3$ (STO) heterostructure superlattice, however, unlike the positive MR for YTO-CTO, a negative MR was observed below this upturn-temperature in SmTO-STO system \cite{SStemmerNC2014}. The observations of low-$\it{T}$ positive MR in case of YTO-CTO rule out Kondo- or weak-localization effects as the mechanisms for which the low-temperature upturn in $\it{R}$-values is observed, as they lead to a negative MR \cite{RCBudhani2014}. The MR of many metals can be analyzed using Kohler's rules, which is based on a semi-classical transport theory using a single species of charge-carriers and isotropic scattering-rates. According to Kohler's rule, the $\it{H}$ dependence of MR at different $\it{T}$ can be related by rescaling $\it{H}$ with by the zero-field ($\it{H}$=0) resistance value ($\it{R}$$_{\rm{0}}$) \cite{ABPippard1989}. As seen in Fig. \ref{ResH}(c), the MR for YTO-CTO superlattices does not follow Kohler's rule as the scaled isothermal MR plots do not overlap. In presence of a single type of charge carrier, as suggested by the Hall data,  and absence of a change in electronic structure with temperature, as suggested by the near-constancy of the charge-carrier concentration with temperature, the violation of Kohler's rule in case of YTO-CTO superlattices suggest towards the interaction of interface charge-carriers with localized spin-moments \cite{JSBrooks1998}. Electron-doped CTO exhibits a negative MR due to incipient localized magnetic moments \cite{HBando2002}, which rules out contribution from any induced magnetization in CTO layers to the MR in YTO-CTO superlattices. The observed MR may, thus, arise from the interaction of the 2DEG interface electrons with the spin moments in YTO layers. For such a scenario, if YTO remains in the FM state, as in bulk, or even in the paramagnetic (PM) state, then it is expected to lead to a negative MR in the YTO-CTO superlattices \cite{JBarnas2001}. Using molecular field approximations, for spin moments aligned in an AFM order, electron-spin scattering mechanism is expected to lead to a positive value of MR \cite{STakada1973}. The same, for FM and PM orders, leads to negative values of MR \cite{STakada1973}. Also, for a metallic layer in proximity to an AFM insulating layer in presence of spin-orbit coupling (which can arise from Rashba spin-orbit coupling in such abrupt thin-film hetero-interfaces) a positive MR (due to Spin-Hall Magnetoresistance [SMR]) is expected to occur for our experimental geometry \cite{SGeprags2018}. Interestingly, as observed in Fig. \ref{ResH}(d), the MR of $\it{m}$=6 superlattice changes sign from positive to negative around the same temperature at which the low-$\it{T}$-upturn in corresponding $\it{R}$-values is observed. Further, no sign change for MR is observed in case of $\it{m}$=3 YTO-CTO superlattice on NGO, suggesting the absence of any magnetic phase transition and consequent upturn in $\it{R}$-values till the lowest temperature. These seem to suggest that the YTO layers, when it undergo an AFM ordering at low-temperatures (as seems to happen for $\it{m}$=6 superlattice on NGO around 40 K and for $\it{m}$=3 on DSO at around 20 K), lead to increased charge-scattering and an increase of low-temperature $\it{R}$ values in the YTO-CTO superlattices.

\indent Further in case of interaction of metallic electrons with AFM-ordered magnetic moments, both molecular field approximation and SMR effect predicts a $\it{H}^{\rm{2}}$-scaling of the MR \cite{STakada1973,SGeprags2018}, as is observed for $\it{m}$=6 superlattice in Fig. \ref{ResH}(a). The larger value of MR for $\it{H}$ applied in-plane than when $\it{H}$ is out-of thin-film plane, as seen in Fig. \ref{ResH}(a), is consistent with the AFM-aligned magnetic moments of YTO lying mostly in the thin-film plane (as usually observed in case of magnetic thin-films due to shape anisotropy effects). The $\it{m}$=3 superlattice sample on NGO, which has only one YTO pseudo-cubic layer away from interfaces (and is possibly too thin to sustain an AFM-order), interestingly, exhibits a different scaling relation of the positive MR with $\it{H}$ (close to $\it{H}^{\rm{1.2}}$), as seen in Fig. \ref{ResH}(b). This scaling relation of MR possibly suggests the presence of increased antiferromagnetic spin-fluctuations even for $\it{H}$=0 and seems to be in consistence with the pronounced non-Fermi-liquid-like charge transport observed till the highest temperature \cite{STakadaPTP1973}.

In summary, we have investigated the transport properties of YTO-CTO superlattices on NGO and DSO substrates containing varying thickness of ultra-thin layers of YTO under different epitaxial strains. The YTO-CTO superlattices host a high-carrier density 2DEG on the CTO side of the heterointerfaces. On increasing $\it{m}$ in the $\it{m}$YTiO$_3$-6CaTiO$_3$ superlattices, a low-temperature upturn in sheet-resistance values and a positive MR below this temperature (which does not follow Kohler's rule), and, which scales as the square of magnetic field (as observed for metallic layers in contact with AFM insulators) are observed. Magnetotransport studies on these superlattices, thus, suggest that this low-temperature upturn in sheet-resistance values corresponds to an antiferromagnetic ordering for the thin YTO layers in the superlattice. This observation is significant as YTO is ferromagnetic in bulk. YTO thin-films, epitaxially grown on LAO (110) substrate were found to remain ferromagnetic instead \cite{CUJung2006}. Further, in view of theoretical predictions, there is a strong likelihood that the obtained antiferromagnetic phase of YTO in YTO-CTO heterostructure is the A-type AFM phase, which is yet to be experimentally realized in any $\it{R}$TiO$_3$ compound. It is, thus, extremely important to probe this emergent magnetic order of YTO in further details using techniques such as polarized neutron reflectometry experiments, which will be the subject of a future study.

We acknowledge the use of PPMS under DST-FIST facility in Department of Physics, IIT Kharagpur for this work. D.C. would like to acknowledge SRIC-IIT Kharagpur (ISIRD grant), SERB, DST, India (funding under project file no. ECR/2016/000019) and BRNS, DAE (funding through sanction number 37(3)/20/23/2016-BRNS) for financial support. J.C. and X.L. were supported by the Gordon and Betty Moore Foundation's EPiQS initiative through Grant no. GBMF4534.

\end{document}